# NP-TCMtarget: a network pharmacology platform for exploring mechanisms of action of Traditional Chinese medicine

Aoyi Wang[a], Yingdong Wang[a], Haoyang Peng[a], Haoran Zhang[a], Caiping Cheng[a], Jinzhong Zhao[a], Wuxia Zhang[a], Jianxin Chen[b,*], Peng Li[a,*]

[a] Shanxi key lab for modernization of TCVM, College of Basic Sciences, Shanxi Agricultural University, Taigu 030801, China

[b] School of Traditional Chinese Medicine, Beijing University of Chinese Medicine, Beijing 100029, China

* Corresponding author.

E-mail addresses: cjx@bucm.edu.cn (J. Chen), lip@sxau.edu.cn (P. Li)

**Abstract:**

The biological targets of traditional Chinese medicine (TCM) are the core effectors mediating the interaction between TCM and the human body. Identification of TCM targets is essential to elucidate the chemical basis and mechanisms of TCM for treating diseases. Given the chemical complexity of TCM, both *in silico* high-throughput drug-target interaction predicting models and biological profile-based methods have been commonly applied for identifying TCM targets based on the structural information of TCM chemical components and biological information, respectively. However, the existing methods lack the integration of TCM chemical and biological information, resulting in difficulty in the systematic discovery of TCM action pathways. To solve this problem, we propose a novel target identification model NP-TCMtarget to explore the TCM target path by combining the overall chemical and biological profiles. First, NP-TCMtarget infers TCM effect targets by calculating associations between drug/disease inducible gene expression profiles and specific gene signatures for 8,233 targets. Then, NP-TCMtarget utilizes a constructed binary classification model to predict binding targets of herbal ingredients. Finally, we can distinguish TCM direct and indirect targets by comparing the effect targets and binding targets to establish the action pathways of herbal components-direct targets-indirect targets by mapping TCM targets in the biological molecular network. We apply NP-TCMtarget to the formula XiaoKeAn to demonstrate the power of revealing the action pathways of herbal formula. We expect that this novel model could provide a systematic framework for exploring the molecular mechanisms of TCM at the target level. NP-TCMtarget is available at http://www.bcxnfz.top/NP-TCMtarget.

## Introduction

Traditional Chinese Medicine (TCM), a treasured human heritage passed down through thousands of years, plays a pivotal role in disease prevention, treatment, and diagnosis. Unlike modern medical theories, the conceptual framework of TCM emphasizes a holistic perspective, employing “syndrome differentiation” by analyzing clinical manifestations to prescribe tailored herbal formulae. These formulae, composed of various medicinal herbs, are designed to address specific disease syndromes. The therapeutic essence of TCM lies in their multifaceted action on biological molecular networks to exert pharmacological effects through a multi-component, multi-target, and multi-pathway approach. This polypharmacological nature is a unique strength of TCM in clinical practice, but, it also underpins the complexity of interactions between TCM and the human body. This complexity hinders the elucidation of the chemical basis and underlying mechanisms of action (MoA) of TCM, thereby impeding its in-depth exploration and advancement (Qiu, 2007; Xu, 2011). Therefore, deciphering the intricate interactions between TCM and the human body has been one of the core challenges in TCM modernization. At the molecular level, TCM can be regarded as a composite of chemical ingredients. After entering the body, these compounds interact with various molecular targets to impact biological functions. The targets are the core effectors mediating the interactions between TCM and the human body. Therefore, the systematic identification of TCM targets is an indispensable step towards elucidating the TCM MoA.

Network pharmacology (NP), as an interdisciplinary of systems biology and pharmacology, intends to systematically interpret the occurrence and development of diseases as well as the MoA of drugs from the perspective of biological networks (Hopkins, 2008). This paradigm naturally fits the systematic characteristics of TCM, giving rise to the methodology of TCM-NP (Li and Zhang, 2013; Luo et al., 2020). In recent years, TCM-NP has been widely applied to explore the chemical basis and molecular mechanisms of TCM drugs, resulting in the establishment of some specialized databases (Li et al., 2015; Zhang et al., 2019), including SymMap (Wu et al., 2019), BATMAN-TCM (Liu et al., 2016), HIT (Ye et al., 2011), TCMID (Xue et al., 2013), TCMSP (Ru et al., 2014), ETCM (Xu et al., 2019), and YaTCM (Li et al., 2018). The main idea of TCM-NP lies in mapping the TCM targets and disease-related genes onto biological networks to build a multilayer TCM component-target-disease network, and systematically analyze the TCM MoA through network analysis. Given the numerous components in TCM, the efficient and rapid identification of the largely unknown biological targets of TCM components is the key problem of TCM-NP. Generally, TCM-NP adopts the efficient and high-throughput virtual prediction technology to capture potential drug-target interactions based on the structural information of TCM chemical components (Zhang et al., 2023). However, these chemo-centric approaches usually tend to predict more binding interactions for existing proteins and ligands and fail to generalize to novel structures (Chatterjee et al., 2023). This poor generalizability should be more prominent in the TCM target prediction, as herbs with multiple components will predict more redundant targets. In addition, the aim of these methods is to predict structurally binding interactions between chemicals and targets, and neglect whether these targets are responsible for the biological effects of the drug, making researchers difficult to concentrate on pharmacologically related targets.

Alternatively, biological profile-based inference methods have the potential to address the limitations of chemo-centric approaches. These methods presume that drugs and their targets should induce similar biological activities, which can serve as the bridge to connect drugs and corresponding targets. Indeed, there have been multitudinous bioactivity profiles in response to chemical and genetic perturbations recorded in public databases like cell growth, gene expression, side effects, and related indications (Schenone et al., 2013). Among them, gene expression data that represent the overall molecular activities of perturbation, are the most suitable adaptors to connect activities and mechanisms of drugs and targets. A prototypical example of this type of study is the Connectivity Map (CMap) which is developed as a public database of >1.3 million gene expression profiles derived from chemical and genetic perturbations of various cell lines, providing opportunities to build functional connections between chemicals and genes at gene expression level (Lamb et al., 2006; Subramanian et al., 2017). The CMap-based target prediction methods have been widely applied in various TCM fields, such as the mechanism study of TCM drugs, synergistic effects within herbal formula, and drug repurposing (Jiang et al., 2021). Moreover, some pharmacotranscriptomic databases have been developed to facilitate the application of gene expression profile-based methods, including HERB, ITCM, and TMNP (Fang et al., 2021; Li et al., 2022; Tian et al., 2023). Biological profile-based methods have the advantage of exploring targets related to the biological effects of TCM (termed effect targets). However, because these methods lack attention to the interactions between TCM components and the effect targets of TCM, it is difficult to determine which targets directly bind with the TCM components to mediate the bioactivity of TCM.

Based on the above analysis, we find that the two strategies discover TCM therapeutic targets from different perspectives. Although they have made much significant contribution to the exploration of the molecular mechanisms of TCM, these methods still have some hierarchical and systematic deficiencies. It is necessary to strictly define the modes of action between drugs and biological targets. To elucidate the complexity of the interactions between drugs and biological targets, we divide the relevant targets for the effects of drugs into two categories: direct and indirect targets. Direct targets are those that directly bind with the drug to produce biological effects. Indirect targets are the subsequent targets influenced by the direct targets in the biological molecular network. Consequently, direct targets should be the intersection of binding targets (targets that structurally bind with the drug) and effect targets (targets that mediate the biological effects of the drug). Indirect targets are the remaining effect targets after excluding the direct targets. As shown in Figure 1, after entering the body, the drug produces biological effects by binding to direct targets or indirectly impacting subsequent signaling pathways through indirect targets. Obviously, exploration of the TCM MoA involves the elucidation of the action pathway of component-direct target-indirect target.

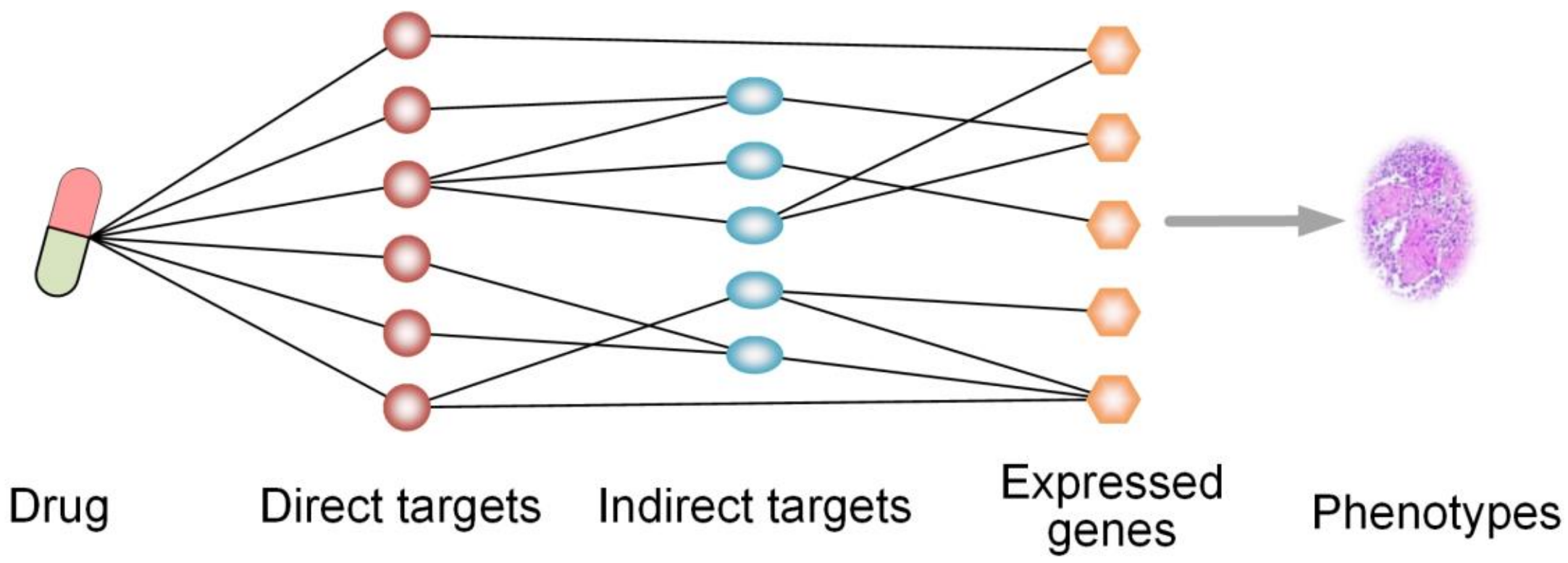

Here, we propose a network pharmacology model NP-TCMtarget that integrates the chemical and biological data of TCM to explore the target path. We first build specific functional gene signatures for 8,233 gene targets. NP-TCMtarget can infer TCM effect targets by calculating associations between drug/disease inducible gene expression profiles and specific target gene signatures. Then, we construct a binary classification model to predict drug-target interactions. NP-TCMtarget can utilize the model to predict the binding targets of herbal ingredients. After obtaining the potential effect targets and binding targets of TCM, we can distinguish direct and indirect targets by comparing the effect targets and binding targets. Finally, we can establish the action pathway of effective component-direct target-indirect target by mapping targets in the biological molecular network (**Figure 2**). We apply NP-TCMtarget to the classic formula XiaoKeAn for treating type 2 diabetes to demonstrate the power of NP-TCMtarget to reveal the action pathways of TCM. In addition, we have embedded NP-TCMtarget in the online platform convenient for users. We believe NP-TCMtarget can provide a holistic framework for studying the molecular mechanisms of TCM.

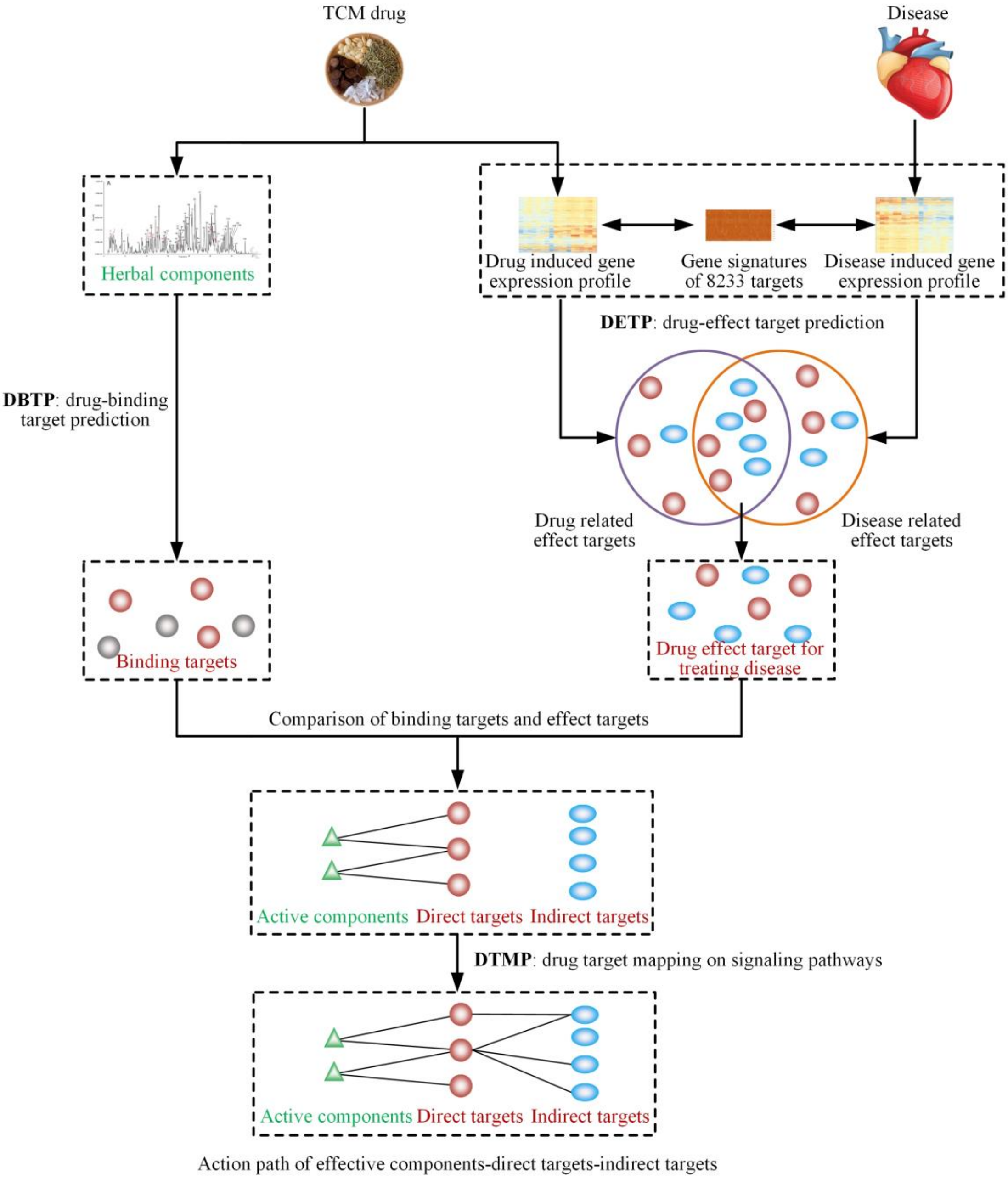

## Materials and methods

### Implementation of NP-TCMtarget

NP-TCMtarget mines effect targets of herbal drugs by calculating associations between drug/disease inducible gene expression profiles and specific target gene signatures embedded in the drug effect target prediction (DETP) module, predicts binding targets of herbal ingredients by the binary classification model in the (Drug binding target prediction) DBTP module, and maps drug targets onto biological pathways by the (drug target mapping on signaling pathways) DTMP module. NP-TCMtarget provides a convenient web interface for users to submit high-throughput transcriptional and chemical data to carry out target path analysis and download data. NP-TCMtarget is freely available at: http://www.bcxnfz.top/NP-TCMtarget. The website was built using the R shiny package and the analysis code was written in R. In addition, we also submitted the R code to GitHub (https://github.com/lipi12q/NP-TCMtarget).

**Gene expression data source and construction of target-specific gene signatures**

Expanded CMap LINCS Resource 2020 (version: beta) is available from LINCS data releases app (https://clue.io/releases/data-dashboard). The L1000 platform carries out a rigorous five-step data processing pipeline to transform raw data from Luminex scanners to replicate-consensus signatures. The L1000 assay directly measures 978 landmark genes and infers additional 11,350 genes. Of the inferred genes, 9,196 are well inferred. Our work only uses the high-fidelity 10,174 genes including 978 measured landmarks and 9,196 well inferred genes. The perturbagens include five types, compounds, shRNA treatments, CRISPR, over-expression (OE) treatments and other treatments. We use LEVEL 5 data for compounds, shRNA, CRISPR and OE, which contains 921,123 signatures, corresponding to 12,487 compounds, 4,345 shRNA, 5,157 CRISPR and 4,040 OE.

We collate gene targets for all chemical perturbagens from the metadata for compounds. Combining target information of compound and genetic perturbagens, all signatures are mapped to 8,233 gene targets. For each target, to get a unique gene signature that accurately measures the on-target activity, we combine gene expression signatures of each target to produce a consensus gene signature by the weighted average algorithm, which calculates a weighted average of the gene expression signatures of each target, with coefficients given by a pairwise Spearman correlation matrix between the expression profiles of all signatures(Smith et al., 2017).

For the expression signature of each target, we sort genes according to their values and selected the top $n$ genes ($t_{up}$) of each gene list and the bottom $n$ ones ($t_{down}$) as the target-specific gene signature ($t_{up}$, $t_{down}$) to represent the transcriptional activity induced by the target.

**Effect target prediction**

Based on the drug or disease-induced gene expression data, we can predict effect targets for drugs or diseases by using the DETP module. The query is a gene list decreasingly ranked by gene metrics, e.g. differential gene expression profiles of the drug/disease. To evaluate the associations between the query gene list and targets, the connections between the query gene list *L* and each target-specific gene signature were calculated by using a modified gene set enrichment approach (Subramanian et al., 2005). We defined the raw similarity score (also named Effect Target Score (ETS) in the present work) as follows:

$$ETS_L^t = \frac{ES_L^{up} - ES_L^{down}}{2}$$

Where $ES_L^{up}$ is the enrichment score of $t_{up}$ for *L* and $ES_L^{down}$ is the enrichment score of $t_{down}$ for *L*. $ETS_L^t$ denotes the effect target score of the gene module pair ($t_{up}$, $t_{down}$) of one target to *L*. *EST* values range between -1 and 1. It measures the degree of similarity between query *L* and target-induced gene expression profiles. It will be positive for targets that are positively related to *L* and negative for those that are inversely similar, and near zero for signatures that are unrelated. A zero value is assigned when both $ES_L^{up}$ and $ES_L^{down}$ are the same sign.

**Estimating significance of effect target score**

We assess the significance of an actual ETS by comparing it with a set of random scores ETSNULL. We generated the random space of query gene lists by randomly permuting genes in *L* for 1000 times, creating a null distribution of $ETS_{NULL}$ for the signature. The nominal P-value for the actual ETS is estimated by using the positive or negative portion of the null distribution corresponding to the sign of the actual ETS, as follows:

$$P = \frac{N(ETS_{NULL} \geq ETS)}{N(CS_{NULL} \geq 0)},\ ETS \geq 0$$

$$P = \frac{N(ETS_{NULL} < ETS)}{N(ETS_{NULL} < 0)},\ ETS < 0$$

**Normalization of effect target score and adjustment for multiple hypothesis testing**

When a query gene list is analyzed against multiple target signatures, we need to adjust the estimated significance level to account for multiple hypothesis testing. Firstly, we normalize the ETS for each target signature to account for the size of the signature, yielding a normalized effect target score (NETS):

$$NETS_L^i = \frac{ETS_L^i}{mean(ETS_{NULL}^i)} \text{ , sign}(ETS_L^i) = \text{sign}(ETS_{NULL}^i)$$

where $ETS_L^i$ is the actual effect target score of the signature i, and the $ETS_{NULL}^i$ is the set of the effect target scores of the signature i against random permutations of the query gene list L with the same signs as $ETS_L^i$. By normalizing ETS, NP-TCMtarget accounts for different sizes in gene signatures and correlations between gene signatures and L. Therefore, NETS can be used to compare analysis results across gene signatures and multiple queries. We then control the proportion of false positives by calculating the false discovery rate (FDR) corresponding to each NETS. The FDR is the estimated probability that a signature with a given NETS represents a false positive finding; it is computed by a ratio of two distributions: (A) the actual effect target score versus the scores for all gene signatures against all permutations of the query gene lists and (B) the actual score versus the scores of all gene signatures against the actual gene lists:

$$A = \frac{N(NETS_{NULL} \geq NETS_L^i)}{N(NETS_{NULL} \geq 0)}, \; NETS_L^i > 0$$

$$A = \frac{N(NETS_{NULL} \leq NETS_L^i)}{N(NETS_{NULL} \leq 0)}, \; NETS_L^i < 0$$

$$B = \frac{N(NETS_{actual} \geq NETS_L^i)}{N(NETS_{actual} \geq 0)}, \; NETS_L^i > 0$$

$$B = \frac{N(NETS_{actual} \leq NETS_L^i)}{N(NETS_{actual} \leq 0)}, \; NETS_L^i < 0$$

$$FDR_L^i = \frac{A}{B}$$

Where $NETS_L^i$ is the actual normalized correlation score of the signature i against L. $NETS_{NULL}$ are the effect target scores for all target signatures against all permutations of the query gene lists. $NCS_{actual}$ are the effect target scores of all target signatures against the actual gene lists.

**Determination of size of target signature**

The size n of $t_{up}$ and $t_{down}$ is heuristically determined as follows. We vary the size $n$ from 100 to 1000 with an interval of 50 and calculated ETS between all target gene expression profiles and gene signatures. It is observed that the overall ETS values (measured by the average of all ETS absolute values at each size) gradually decrease with the increase of the size of signatures. Moreover, we observe that the extent of decrease of ETS values is gradually slowing with the size increases, suggesting that with the size of gene module pair increase to some extent, its influence on the overall ETS values can be minimal. We quantify the decrease extent of ETS by the decrease rate (DR):

$$DR = \frac{mean(|ETS_{i+50}| - |ETS_i|)}{mean(|ETS_i|)}$$

Finally, we choose n = 350 as a good compromise, as the decrease rate of ETS at this size is small enough (DR = ~0.01, Supplementary Figure 1).

**Drug-target interaction dataset**

Drug-target interaction datasets are collected from DrugBank, which is the most widely used drug information resource. We download the latest dataset from the DrugBank website (https://go.drugbank.com/releases/latest). Finally, 19,891 interactions with 3,894 drugs and their 6,898 protein targets are retained. All these data are used to build the drug-binding target prediction model. These data can be available at https://github.com/lipi12q/NP-TCMtarget.

**Drug binding target prediction by DeepPurpose**

DeepPurpose is a deep learning library for encoding and downstream prediction of compound-protein interactions, in which models are usually considered as state-of-the-art methods for the prediction of drug-target interactions(Huang et al., 2021). In the present work, the model is implemented in the DBTP module. DBTP takes the simplified molecular-input line entry system (SMILES) string of drugs and amino acid sequences of proteins as input, which are both encoded with the convolutional neural network (CNN). Then the drug and protein embeddings are fed into a multilayer perceptron (MLP) decoder to generate binary outputs indicating the probability that a compound binds to a protein (also named drug binding score). During training, the learning rate was set to 0.0008, the batch size was 256, and the overall training was conducted for 100 epochs. DBTP uses Binary Cross Entropy as the loss function and Area Under the Receiver Operating Characteristics (AUROC) and Area Under Precision-Recall (AUPRC) as performance metrics.

## Results

### Scheme of drug target path prediction

NP-TCMtarget includes three main modules: **d**rug **e**ffect **t**arget **p**rediction based on the drug-induced gene expression data (DETP), **d**rug **b**inding **t**arget **p**rediction using the deep learning model (DBTP), and **d**rug **t**arget **m**apping on signaling **p**athways (DTMP). DETP has built a library of target-specific signatures derived from four types of gene expression data induced by four types of perturbations: compound, shRNA, CRISPR and OE. Then using gene expression data induced by drugs or diseases as a query, DETP can evaluate the effects of drugs on targets by measuring the correlation between the drug/disease inducible gene expression profiles and target signatures using the association algorithm (See methods). DBTP has constructed a deep learning drug-target interaction prediction model, which can be utilized to predict the binding probability between chemical ingredients of herbs and targets (See methods). DTMP enriches drug targets into biological pathways and distinguishes direct and indirect targets within the pathways, allowing for the exploration of MoA of herbal medicine.

### Application of NP-TCMtarget

The core objective of NP-TCMtarget is to explore the molecular mechanisms of herbal medicine for treating diseases based on high-throughput transcriptional and chemical data of herbal medicine. The primary input of NP-TCMtarget has two parts: herbal drug or disease inducible gene expression profiles and chemical ingredients of the corresponding herbs (**Figure 2 and 3**). In the calculation, the query gene expression data should be ranked lists in descending order of some differential metrics (e.g., the Signal2Noise or fold change values between drug-treated vs. control samples). DETP estimates the impact of the drug or disease on each target by the resultant normalized effect target score (NETS) and the corresponding false discovery rate (FDR), which are calculated based on the correlation between the gene expression profiles and the catalog of target signatures. Generally, those targets with FDR $\leq 0.05$ are regarded as potential effect targets of the query drug or disease. More importantly, the sign (positive or negative) of NETS values reflects different action modes on the target. If the drug and disease have opposite NETS values for the same targets, the drug has the potential to treat the disease by these targets. Therefore, by comparing the effect targets of the drug and the disease, we can obtain the therapeutic targets that may play beneficial effects in the drug treatment for the disease.

In the NP-TCMtarget platform, the structural information of herbal ingredients is represented by canonical SMILES and inputted into the DBTP module (**Figure 2 and 3**). DBTP uses the embedded drug-target interaction prediction model to predict the binding targets of chemical ingredients. The predicted targets with binding scores above a threshold value are regarded as potential binding targets of herbal ingredients. We can compare effect targets to binding targets of herbs and obtain direct targets of herbs which are the intersection of effect targets and binding targets. The remaining effect targets are seen as indirect targets of herbs.

DTMP enriches the effect targets into their corresponding biological signaling pathways and distinguishes between direct and indirect targets within the pathways. Various biological networks such as protein-protein interaction networks, gene co-expression networks, gene function networks, and molecular signaling pathway networks can be utilized to map all targets to their respective networks, allowing for the exploration of signaling pathways of herbal targets and laying the foundation for the study of the molecular mechanism. As various known biological networks are applied in different fields without a clear distinction of superiority or inferiority, the choice of the appropriate molecular network can be flexible according to one's own research background and research objectives. The default option is to use the widely utilized KEGG molecular pathway network in gene enrichment analysis. In the NP-TCMtarget website, the above procedures are performed automatically on the server and all results can be easily reviewed and downloaded by users on the "Results" page (**Figure 3**).

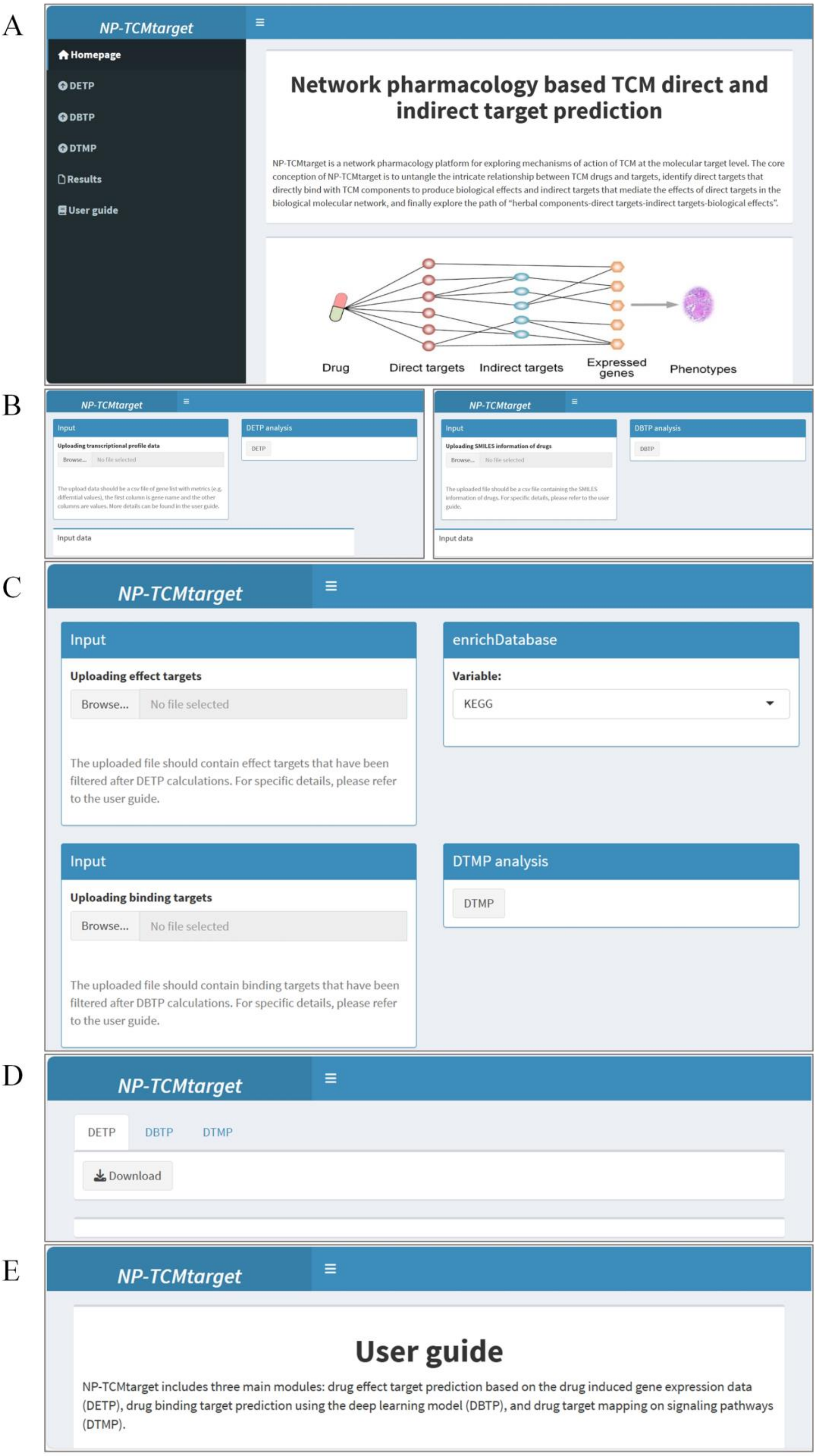
A
NP-TCMtarget
Homepage
DETP
DBTP
DTMP
Results
User guide
Network pharmacology based TCM direct and indirect target prediction
NP-TCMtarget is a network pharmacology platform for exploring mechanisms of action of TCM at the molecular target level. The core conception of NP-TCMtarget is to untangle the intricate relationship between TCM drugs and targets, identify direct targets that directly bind with TCM components to produce biological effects and indirect targets that mediate the effects of direct targets in the biological molecular network, and finally explore the path of "herbal components-direct targets-indirect targets-biological effects".
Drug
Direct targets
Indirect targets
Expressed genes
Phenotypes
B
NP-TCMtarget
Input
Uploading transcriptional profile data
Browse...
No file selected
DETP analysis
DETP
Input data
NP-TCMtarget
Input
Uploading SMILES information of drugs
Browse...
No file selected
DBTP analysis
DBTP
Input data
C
NP-TCMtarget
Input
Uploading effect targets
Browse...
No file selected
The uploaded file should contain effect targets that have been filtered after DETP calculations. For specific details, please refer to the user guide.
enrichDatabase
Variable:
KEGG
Input
Uploading binding targets
Browse...
No file selected
The uploaded file should contain binding targets that have been filtered after DBTP calculations. For specific details, please refer to the user guide.
DTMP analysis
DTMP
D
NP-TCMtarget
DETP
DBTP
DTMP
Download
E
NP-TCMtarget
User guide
NP-TCMtarget includes three main modules: drug effect target prediction based on the drug induced gene expression data (DETP), drug binding target prediction using the deep learning model (DBTP), and drug target mapping on signaling pathways (DTMP).

## Use Case

### Exploring molecular mechanisms of XiaoKeAn against type 2 diabetes

Type 2 diabetes is a common chronic disease with persistently elevated blood glucose concentration and the pathophysiology is characterized by insulin resistance and initial hyperinsulinemia (Ahmad et al., 2022). XiaoKeAn is a traditional Chinese medicine preparation specifically designed to manage type 2 diabetes. Classified as a national Class III new drug in China, XiaoKeAn demonstrates significant efficacy in reducing blood sugar levels and alleviating associated symptoms. However, despite its widespread use, there remains a dearth of research exploring the molecular mechanisms underlying its therapeutic effects in diabetes treatment. This section aims to employ NP-TCMtarget, leveraging transcriptomic data and chemical information of XiaoKeAn, to elucidate the mechanism of XiaoKeAn in treating type 2 diabetes at the molecular level.

Following the instruction of NP-TCMtarget, we first collect the transcriptional data from the microarray experiment deposited in GEO (GSE accession: GSE62087). The experiment includes three groups: the normal control group (C57BL/6J mice), the disease model group (KKAy mice) and the corresponding XiaoKeAn treatment group. The differential gene profiles induced by the diabetes model and XiaoKeAn treatment are calculated and submitted to the DETP module to predict effect targets. We find that the disease model and XiaoKeAn treatment are significantly associated with 1,612 and 2,157 effect targets among all 8,233 targets, respectively (Supplementary Table 1; FDR $\leq$ 0.05; **Figure 4A**). We observe a negative correlation between the disease model and XiaoKeAn treatment that is quantified by the Pearson correlation of NETS values for all targets (PCC = -0.50; **Figure 4B**), indicating that XiaoKeAn can reverse the impact of diabetes. The absolute value of NETS reflects the magnitude of the drug/disease’s effects on a target, with a higher absolute value indicating a stronger impact. If the NETS values for the drug and disease on a specific target are opposite, it suggests opposing effects of the drug and disease on that target. By comparing the NETS values of the effect targets that are significantly associated with the diabetes model and XiaoKeAn treatment, we identify 734 targets that were significantly affected by the diabetes model but could be reversed by the XiaoKeAn treatment (indicated by opposing signs in NETS values; FDR ≤ 0.05; Supplementary table 1). More importantly, we observe that there is a strong negative correlation between the disease model and XiaoKeAn treatment if we only focus on these 734 targets (PCC = -0.98; **Figure 4C**). These results suggest that the 734 targets can be potential therapeutic targets for XiaoKeAn in the treatment of diabetes mellitus.

In addition, a total of 820 chemical components are gathered from the SymMap database for eight herbs in the XiaoKeAn formula (Supplementary Table 2). Utilizing the SMILE information of all these components, the DBTP module is employed to predict 5,030 binding targets of XiaoKeAn ingredients (Supplementary Table 3; Binding score ≥ 0.90). By comparing these 5,030 binding targets of XiaoKeAn and its 734 effect targets for treating diabetes, we obtain 421 direct targets and 313 indirect targets of XiaoKeAn for treating diabetes (Supplementary Table 4; **Figure 4A**). These targets can further be mapped to biological pathways to elucidate the molecular mechanisms of XiaoKeAn in diabetes treatment.

Through enrichment analysis in the DTMP module, the 734 potential therapeutic targets are significantly enriched in 46 functional pathways (FDR ≤ 0.05, **Figure 4A and Table 1**). Detailed examination of these pathways reveals three primary targeted mechanisms of XiaoKeAn in diabetes treatment: (1) regulation of glucose metabolism, such as "Carbohydrate digestion and absorption", "insulin signaling pathway" and "insulin secretion"; (2) regulation of endocrine-related functions, such as "Aldosterone-regulated sodium reabsorption", "Thyroid hormone signaling pathway", "Neurotrophin signaling pathway", "Proximal tubule bicarbonate reclamation", "Bile secretion", "Mineral absorption", "Thyroid hormone synthesis" and "Endocrine and other factor-regulated calcium reabsorption", and (3) regulation of inflammatory immune responses, such as "AGE-RAGE signaling in diabetic complications", "Human T-cell leukemia virus 1 infection", "Influenza A", "C-type lectin receptor signaling pathway", "TNF signaling pathway", "B cell receptor signaling pathway" and "JAK-STAT signaling". These findings indicate that XiaoKeAn regulates multiple functions to treat Type 2 diabetes, aligning with previous reports on its effectiveness against this condition(Yang et al., 2015a; Yang et al., 2015b).

DTMP also indicates the direct and indirect targets in each pathway, in which we can inspect the detailed target route of XiaoKeAn. For example, the DTMP analysis indicates that "AGE-RAGE signaling pathway in diabetic complications" is one of all 46 pathways significantly related to type 2 diabetes (FDR = 0.04). Actually, it has been well known that the AGE-RAGE signaling pathway plays a crucial role in the development of diabetic complications. AGEs (advanced glycation end products) are accumulated under hyperglycemic conditions and activate the receptor of AGEs (RAGE). This activation triggers a cascade of adverse responses, including inflammation, oxidative stress, endothelial cell dysfunction with apoptosis, and eventually diabetic complications, including retinopathy, neuropathy, cardiomyopathy, microvascular complications, and nephropathy(Egaña-Gorroño et al., 2020; Rungratanawanich et al., 2021). Therefore, drug development and therapeutic strategies targeting the AGE-RAGE signaling pathway may offer new directions for preventing and treating diabetic complications. The DTMP analysis displays that XiaoKeAn can regulate this pathway by 6 direct targets including AKT2, COL4A4, PIK3CA, PIK3CB, SMAD4 and STAT1, and 7 indirect targets including EDN1, HRAS, ICAM1, KRAS, NOX1, VEGFB and VEGFD (**Table 1** and **Figure 4D**). Moreover, we can find 631 chemical components that hit the 6 direct targets from the results of the DBTP analysis. Based on these results, we can explore the molecular mechanisms of XiaoKeAn for regulating the AGE-RAGE signaling pathway from the route of chemical components-direct targets-indirect targets (**Figure 4D**).

This section displays that NP-TCMtarget can effectively reveal the chemical basis and relevant mechanisms of XiaoKeAn in the treatment of type 2 diabetes from the target level, and provide an example for the exploration of the mechanisms of action of herbal formula for treating complex diseases.

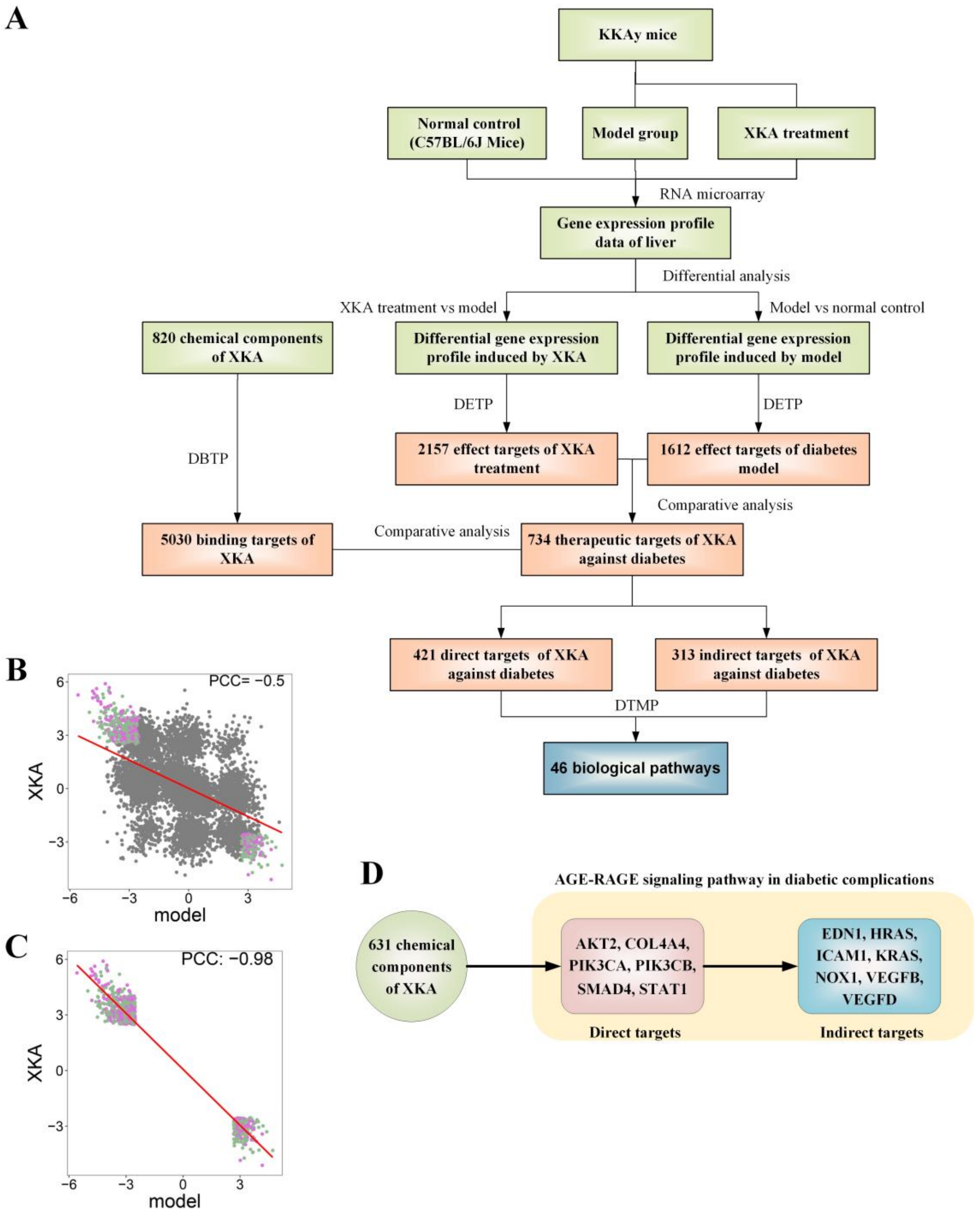

## Discussion

Elucidating the intricate interactions between herbs and the human body remains a pivotal issue in the TCM field. Herbal targets, acting as crucial mediators in these interactions, are essential for elucidating the underlying mechanisms of these interactions. We here systematically delineate the sequential pipeline of "herbal components-direct targets-indirect targets-biological effects" through which TCM drugs interact with the human body. Based on this framework, we introduce the NP-TCMtarget strategy to facilitate the identification of TCM targets and further enable the traversal of the path of "herbal component-direct target-indirect target-biological effect" by integrating the chemical and biological information of TCM cohesively.

The core conception of NP-TCMtarget is to untangle the intricate relationship between drugs and biological targets and identify direct targets that directly bind with the drug to produce biological effects and indirect targets that are subsequently influenced by the direct targets in the biological molecular network (**Figure 1**). To achieve this goal, NP-TCMtarget firstly predicts effect targets that mediate the biological effects of the drug and should be the union of direct and indirect targets based on the transcriptional profiles of drugs. Then NP-TCMtarget predicts binding targets that structurally bind with the drug. By comparing effect targets and binding targets, we can get direct targets and indirect targets of the drug, and further elucidate the action path of effective component-direct target-indirect target.

The idea of predicting drug effect targets in the DETP module of NP-TCMtarget is derived from the CMap-based methods, which consider two perturbagens with similar gene expression profiles that have similar molecular targets. However, although CMap dataset has curated numerous chemical and genetic perturbations with gene expression profiles (https://clue.io/) (Subramanian et al., 2017), the knowledge of the landmarked perturbations is still a severe constraint for the application of the CMap-based methods. NP-TCMtarget solves this problem by manually constructing gene signatures of each target from prior knowledge. The targets of a query drug can be inferred by directly comparing its inducible gene expression profiles with the target-specific gene signatures.

The prediction of drug binding targets embedded in the DBTP module of NP-TCMtarget is based on the chemogenomic information to formalize binding interactions between chemicals and targets. Chemogenomics-based methods utilize both machine learning and artificial intelligence algorithms to learn molecular representations and capture the complex nonlinear relationships between chemicals and target proteins (Chen et al., 2018). NP-TCMtarget uses the DeepPurpose framework which contains known state-of-the-art deep learning models that have been developed and leveraged for the prediction of drug-target interactions (Huang et al., 2021). Based on the DeepPurpose, NP-TCMtarget can effectively predict binding targets for herbal ingredients. More importantly, by comparing effect targets and binding targets, we can get direct targets of herbal ingredients. The direct target space is responsible for the biological function of the drug and is greatly compressed compared with the predicted binding target set which is usually redundant.

In summary, NP-TCMtarget combines TCM transcriptomic and chemical data to explore the underlying TCM MoA for treating diseases at the molecular target level. The ability of NP-TCMtarget can be impacted by some limitations. The result of predicting effect targets is affected by the quantity and quality of the target-specific signatures. We have created molecular signatures for 8,233 gene targets by their inducible gene expression data in the CMap dataset. However, the gene expression data can be affected by many factors, including cell types, drug doses, genetic types, and time of duration. This data heterogeneity significantly impacts the quantity of resulting gene signatures for each target. In addition, the scale of 8,233 targets is still inadequate relative to the overall genome and can be improved. To facilitate users to flexibly apply the algorithm for the customized transcriptomic data and gene signatures, except for the online NP-TCMtarget website, we have submitted the R code to GitHub (https://github.com/lipi12q/ NP-TCMtarget).

**Data availability**

The data used in the present work are available at https://github.com/lipi12q/NP-TCMtarget. All data used in this work are from public resources. The CMap source can be found at https://clue.io/releases/data-dashboard. The Drugbank source can be found at https://go.drugbank.com/releases/latest.

**Code availability**

The codes of this work are freely available at GitHub (https://github.com/lipi12q/NP-TCMtarget).

**Funding**

This research was supported by the National Natural Science Fund of China (No. 82274363), the Fundamental Research Program of Shanxi Province (No. 20210302124129), the Distinguished and Excellent Young Scholars Cultivation Project of Shanxi Agricultural University (No. 2022YQPYGC09), and the Postgraduate Research Innovation Project of Shanxi Province (No. 2023KY322).

**References**

Ahmad, E., Lim, S., Lamptey, R., Webb, D.R., Davies, M.J., 2022. Type 2 diabetes. Lancet (London, England) 400(10365), 1803-1820. https://doi.org/10.1016/s0140-6736(22)01655-5.

Chatterjee, A., Walters, R., Shafi, Z., Ahmed, O.S., Sebek, M., Gysi, D., Yu, R., Eliassi-Rad, T., Barabási, A.L., Menichetti, G., 2023. Improving the generalizability of protein-ligand binding predictions with AI-Bind. Nature communications 14(1), 1989. https://doi.org/10.1038/s41467-023-37572-z.

Chen, H., Engkvist, O., Wang, Y., Olivecrona, M., Blaschke, T., 2018. The rise of deep learning in drug discovery. Drug discovery today 23(6), 1241-1250. https://doi.org/10.1016/j.drudis.2018.01.039.

Egaña-Gorroño, L., López-Díez, R., Yepuri, G., Ramirez, L.S., Reverdatto, S., Gugger, P.F., Shekhtman, A., Ramasamy, R., Schmidt, A.M., 2020. Receptor for Advanced Glycation End Products (RAGE) and Mechanisms and Therapeutic Opportunities in Diabetes and Cardiovascular Disease: Insights From Human Subjects and Animal Models. Frontiers in cardiovascular medicine 7, 37. https://doi.org/10.3389/fcvm.2020.00037.

Fang, S., Dong, L., Liu, L., Guo, J., Zhao, L., Zhang, J., Bu, D., Liu, X., Huo, P., Cao, W., Dong, Q., Wu, J., Zeng, X., Wu, Y., Zhao, Y., 2021. HERB: a high-throughput experiment- and reference-guided database of traditional Chinese medicine. Nucleic acids research 49(D1), D1197-d1206. https://doi.org/10.1093/nar/gkaa1063.

Hopkins, A.L., 2008. Network pharmacology: the next paradigm in drug discovery. Nature chemical biology 4(11), 682-690. https://doi.org/10.1038/nchembio.118.

Huang, K., Fu, T., Glass, L.M., Zitnik, M., Xiao, C., Sun, J., 2021. DeepPurpose: a deep learning library for drug-target interaction prediction. Bioinformatics (Oxford, England) 36(22-23), 5545-5547. https://doi.org/10.1093/bioinformatics/btaa1005.

Jiang, H., Hu, C., Chen, M., 2021. The Advantages of Connectivity Map Applied in Traditional Chinese Medicine. Frontiers in pharmacology 12, 474267.

https://doi.org/10.3389/fphar.2021.474267.

Lamb, J., Crawford, E.D., Peck, D., Modell, J.W., Blat, I.C., Wrobel, M.J., Lerner, J., Brunet, J.P., Subramanian, A., Ross, K.N., Reich, M., Hieronymus, H., Wei, G., Armstrong, S.A., Haggarty, S.J., Clemons, P.A., Wei, R., Carr, S.A., Lander, E.S., Golub, T.R., 2006. The Connectivity Map: using gene-expression signatures to connect small molecules, genes, and disease. Science (New York, N.Y.) 313(5795), 1929-1935. https://doi.org/10.1126/science.1132939.

Li, B., Ma, C., Zhao, X., Hu, Z., Du, T., Xu, X., Wang, Z., Lin, J., 2018. YaTCM: Yet another Traditional Chinese Medicine Database for Drug Discovery. Computational and structural biotechnology journal 16, 600-610. https://doi.org/10.1016/j.csbj.2018.11.002.

Li, P., Fu, Y., Wang, Y., 2015. Network based approach to drug discovery: a mini review. Mini reviews in medicinal chemistry 15(8), 687-695. https://doi.org/10.2174/1389557515666150219143933.

Li, P., Zhang, H., Zhang, W., Zhang, Y., Zhan, L., Wang, N., Chen, C., Fu, B., Zhao, J., Zhou, X., Guo, S., Chen, J., 2022. TMNP: a transcriptome-based multi-scale network pharmacology platform for herbal medicine. Briefings in bioinformatics 23(1). https://doi.org/10.1093/bib/bbab542.

Li, S., Zhang, B., 2013. Traditional Chinese medicine network pharmacology: theory, methodology and application. Chinese journal of natural medicines 11(2), 110-120. https://doi.org/10.1016/s1875-5364(13)60037-0.

Liu, Z., Guo, F., Wang, Y., Li, C., Zhang, X., Li, H., Diao, L., Gu, J., Wang, W., Li, D., He, F., 2016. BATMAN-TCM: a Bioinformatics Analysis Tool for Molecular mechANism of Traditional Chinese Medicine. Scientific reports 6, 21146. https://doi.org/10.1038/srep21146.

Luo, T.T., Lu, Y., Yan, S.K., Xiao, X., Rong, X.L., Guo, J., 2020. Network Pharmacology in Research of Chinese Medicine Formula: Methodology, Application and Prospective. Chinese journal of integrative medicine 26(1), 72-80. https://doi.org/10.1007/s11655-019-3064-0.

Qiu, J., 2007. Traditional medicine: a culture in the balance. Nature 448(7150), 126-

128. https://doi.org/10.1038/448126a.

Ru, J., Li, P., Wang, J., Zhou, W., Li, B., Huang, C., Li, P., Guo, Z., Tao, W., Yang, Y., Xu, X., Li, Y., Wang, Y., Yang, L., 2014. TCMSP: a database of systems pharmacology for drug discovery from herbal medicines. Journal of cheminformatics 6, 13. https://doi.org/10.1186/1758-2946-6-13.

Rungratanawanich, W., Qu, Y., Wang, X., Essa, M.M., Song, B.J., 2021. Advanced glycation end products (AGEs) and other adducts in aging-related diseases and alcohol-mediated tissue injury. Experimental & molecular medicine 53(2), 168-188. https://doi.org/10.1038/s12276-021-00561-7.

Schenone, M., Dančík, V., Wagner, B.K., Clemons, P.A., 2013. Target identification and mechanism of action in chemical biology and drug discovery. Nature chemical biology 9(4), 232-240. https://doi.org/10.1038/nchembio.1199.

Smith, I., Greenside, P.G., Natoli, T., Lahr, D.L., Wadden, D., Tirosh, I., Narayan, R., Root, D.E., Golub, T.R., Subramanian, A., Doench, J.G., 2017. Evaluation of RNAi and CRISPR technologies by large-scale gene expression profiling in the Connectivity Map. PLoS biology 15(11), e2003213. https://doi.org/10.1371/journal.pbio.2003213.

Subramanian, A., Narayan, R., Corsello, S.M., Peck, D.D., Natoli, T.E., Lu, X., Gould, J., Davis, J.F., Tubelli, A.A., Asiedu, J.K., Lahr, D.L., Hirschman, J.E., Liu, Z., Donahue, M., Julian, B., Khan, M., Wadden, D., Smith, I.C., Lam, D., Liberzon, A., Toder, C., Bagul, M., Orzechowski, M., Enache, O.M., Piccioni, F., Johnson, S.A., Lyons, N.J., Berger, A.H., Shamji, A.F., Brooks, A.N., Vrcic, A., Flynn, C., Rosains, J., Takeda, D.Y., Hu, R., Davison, D., Lamb, J., Ardlie, K., Hogstrom, L., Greenside, P., Gray, N.S., Clemons, P.A., Silver, S., Wu, X., Zhao, W.N., Read-Button, W., Wu, X., Haggarty, S.J., Ronco, L.V., Boehm, J.S., Schreiber, S.L., Doench, J.G., Bittker, J.A., Root, D.E., Wong, B., Golub, T.R., 2017. A Next Generation Connectivity Map: L1000 Platform and the First 1,000,000 Profiles. Cell 171(6), 1437-1452.e1417. https://doi.org/10.1016/j.cell.2017.10.049.

Subramanian, A., Tamayo, P., Mootha, V.K., Mukherjee, S., Ebert, B.L., Gillette, M.A., Paulovich, A., Pomeroy, S.L., Golub, T.R., Lander, E.S., Mesirov, J.P., 2005. Gene set enrichment analysis: a knowledge-based approach for interpreting genome-wide expression profiles. Proceedings of the National Academy of Sciences of the United States of America 102(43), 15545-15550. https://doi.org/10.1073/pnas.0506580102.

Tian, S., Zhang, J., Yuan, S., Wang, Q., Lv, C., Wang, J., Fang, J., Fu, L., Yang, J., Zu, X., Zhao, J., Zhang, W., 2023. Exploring pharmacological active ingredients of traditional Chinese medicine by pharmacotranscriptomic map in ITCM. Briefings in bioinformatics 24(2). https://doi.org/10.1093/bib/bbad027.

Wu, Y., Zhang, F., Yang, K., Fang, S., Bu, D., Li, H., Sun, L., Hu, H., Gao, K., Wang, W., Zhou, X., Zhao, Y., Chen, J., 2019. SymMap: an integrative database of traditional Chinese medicine enhanced by symptom mapping. Nucleic acids research 47(D1), D1110-d1117. https://doi.org/10.1093/nar/gky1021.

Xu, H.Y., Zhang, Y.Q., Liu, Z.M., Chen, T., Lv, C.Y., Tang, S.H., Zhang, X.B., Zhang, W., Li, Z.Y., Zhou, R.R., Yang, H.J., Wang, X.J., Huang, L.Q., 2019. ETCM: an encyclopaedia of traditional Chinese medicine. Nucleic acids research 47(D1), D976-d982. https://doi.org/10.1093/nar/gky987.

Xu, Z., 2011. Modernization: One step at a time. Nature 480(7378), S90-92. https://doi.org/10.1038/480S90a.

Xue, R., Fang, Z., Zhang, M., Yi, Z., Wen, C., Shi, T., 2013. TCMID: Traditional Chinese Medicine integrative database for herb molecular mechanism analysis. Nucleic acids research 41(Database issue), D1089-1095. https://doi.org/10.1093/nar/gks1100.

Yang, Z., Wang, L., Zhang, F., Li, Z., 2015a. Evaluating the antidiabetic effects of Chinese herbal medicine: Xiao-Ke-An in 3T3-L1 cells and KKAy mice using both conventional and holistic omics approaches. BMC complementary and alternative medicine 15, 272. https://doi.org/10.1186/s12906-015-0785-2.

Yang, Z.Z., Liu, W., Zhang, F., Li, Z., Cheng, Y.Y., 2015b. Deciphering the therapeutic mechanisms of Xiao-Ke-An in treatment of type 2 diabetes in mice by a Fangjiomics approach. Acta pharmacologica Sinica 36(6), 699-707. https://doi.org/10.1038/aps.2014.138.

Ye, H., Ye, L., Kang, H., Zhang, D., Tao, L., Tang, K., Liu, X., Zhu, R., Liu, Q., Chen, Y.Z., Li, Y., Cao, Z., 2011. HIT: linking herbal active ingredients to targets. Nucleic acids research 39(Database issue), D1055-1059. https://doi.org/10.1093/nar/gkq1165.

Zhang, P., Zhang, D., Zhou, W., Wang, L., Wang, B., Zhang, T., Li, S., 2023. Network pharmacology: towards the artificial intelligence-based precision traditional Chinese medicine. Briefings in bioinformatics 25(1). https://doi.org/10.1093/bib/bbad518.

Zhang, R., Zhu, X., Bai, H., Ning, K., 2019. Network Pharmacology Databases for Traditional Chinese Medicine: Review and Assessment. Frontiers in pharmacology 10, 123. https://doi.org/10.3389/fphar.2019.00123.

**Table 1 Action pathways of XKA in treating type 2 diabetes**

| Pathway ID | description | FDR | Direct target | Indirect target |
|---|---|---|---|---|
| hsa04973 | Carbohydrate digestion and absorption | 0.00 | AKT2;ATP1A2;ATP1A3;ATP1A4;ATP1B3;HK2;HK3;PIK3CA;PIK3CB;SLC5A1 | ATP1B1;ATP1B4;FXYD2 |
| hsa04960 | Aldosterone-regulated sodium reabsorption | 0.00 | ATP1A2;ATP1A3;ATP1A4;ATP1B3;PIK3CA;PIK3CB | ATP1B1;ATP1B4;FXYD2;HSD11B2;KRAS |
| hsa04012 | ErbB signaling pathway | 0.00 | AKT2;CAMK2G;ERBB3;MTOR;NRG1;PAK4;PIK3CA;PIK3CB | ABL2;CAMK2B;CRKL;HRAS;KRAS;MAP2K4;MYC;SHC1 |
| hsa04218 | Cellular senescence | 0.00 | AKT2;ATR;CCNA1;CDC25A;E2F3;E2F5;MTOR;MYBL2;PIK3CA;PIK3CB;PPP1CA;PPP1CC;RAD50;RASSF5;SLC25A6 | CALM2;HIPK1;HRAS;KRAS;MAP2K3;MYC;PPP3R1;VDAC2 |
| hsa04261 | Adrenergic signaling in cardiomyocytes | 0.00 | AKT2;ATP1A2;ATP1A3;ATP1A4;ATP1B3;ATP2A2;CAMK2G;PIK3CG;PPP1CA;PPP1CC;PPP2R3A | ATP1B1;ATP1B4;CALM2;CAMK2B;FXYD2;PPP2CA;PRKACB;RPS6KA5;TPM3;TPM4 |
| hsa04919 | Thyroid hormone signaling pathway | 0.00 | AKT2;ATP1A2;ATP1A3;ATP1A4;ATP1B3;ATP2A2;MED12;MTOR;PIK3CA;PIK3CB;STAT1 | ATP1B1;ATP1B4;FXYD2;HRAS;KRAS;MYC;PRKACB |
| hsa04722 | Neurotrophin signaling pathway | 0.01 | AKT2;CAMK2G;FRS2;MAGED1;MAP3K5;NTRK2;PIK3CA;PIK3CB;RPS6KA1 | CALM2;CAMK2B;CRKL;HRAS;KRAS;PSEN1;RHOA;RPS6KA5;SHC1 |
| hsa05205 | Proteoglycans in cancer | 0.01 | AKT2;CAMK2G;DROSHA;ERBB3;FRS2;FZD7;IGF1R;IQGAP1;ITGA2;MTOR;PIK3CA;PIK3CB;PPP1CA;PPP1CC;WNT1;WNT3 | CAMK2B;HRAS;KRAS;MYC;NANOG;PRKACB;RHOA;TFAP4;TWIST1 |
| hsa04024 | cAMP signaling pathway | 0.01 | ABCC4;AKT2;ATP1A2;ATP1A3;ATP1A4;ATP1B3;ATP2A2;CAMK2G;GLI1;GLI3;HCAR2;PIK3CA;PIK3CB;PPP1CA;PPP1CC;VAV3 | ATP1B1;ATP1B4;CALM2;CAMK2B;FXYD2;GHSR;GLP1R;PRKACB;RHOA |
| hsa04964 | Proximal tubule bicarbonate reclamation | 0.01 | ATP1A2;ATP1A3;ATP1A4;ATP1B3 | ATP1B1;ATP1B4;FXYD2 |
| hsa05230 | Central carbon metabolism in cancer | 0.01 | AKT2;G6PD;HK2;HK3;MTOR;PDHA1;PIK3CA;PIK3CB;SIRT6 | HRAS;KRAS;MYC |
| hsa05166 | Human T-cell leukemia virus 1 infection | 0.01 | AKT2;APC;ATR;E2F3;FZD7;MYBL2;PIK3CA;PIK3CB;SLC25A6;SMAD4;WNT1;WNT3 | CANX;ELK4;HLA-DMB;HLA-DQA2;HRAS;ICAM1;IL2RA;IL2RG;KRAS;MAP2K4;MYC;NFYB;PCNA;POLE3;PPP3R1;PRKACB;VDAC2 |
| hsa05164 | Influenza A | 0.01 | ADAR;AKT2;CIITA;FURIN;IFN | CPSF4;HLA-DMB;HLA- |
| | | | AR2;IRF3;NLRP3;PIK3CA;PIK3CB;PRSS2;SLC25A6;STAT1 | DQA2;ICAM1;IL18;MAP2K3;MAP2K4;RNASEL;TLR7;TNFRSF10B |
| hsa04911 | Insulin secretion | 0.01 | ATP1A2;ATP1A3;ATP1A4;ATP1B3;CAMK2G;PCLO | ATP1B1;ATP1B4;CAMK2B;FFAR1;FXYD2;GLP1R;KCNMB2;PRKACB |
| hsa03010 | Ribosome | 0.01 | RPL3 | RPL10L;RPL11;RPL13A;RPL15;RPL19;RPL23;RPL23A;RPL26L1;RPL37;RPL37A;RPL7;RPL8;RPS15A;RPS2;RPS3;RPS9;RSL24D1 |
| hsa04976 | Bile secretion | 0.01 | ABCC4;ATP1A2;ATP1A3;ATP1A4;ATP1B3;SLC5A1 | AQP9;ATP1B1;ATP1B4;FXYD2;NR0B2;PRKACB |
| hsa05214 | Glioma | 0.01 | AKT2;CAMK2G;E2F3;IGF1R;MTOR;PIK3CA;PIK3CB | CALM2;CAMK2B;HRAS;KRAS;SHC1 |
| hsa04213 | Longevity regulating pathway | 0.01 | AKT2;CAT;IGF1R;MTOR;PIK3CA;PIK3CB | HRAS;KRAS;PRKACB;SOD1;SOD2 |
| hsa04022 | cGMP-PKG signaling pathway | 0.01 | ADRA2A;AKT2;ATP1A2;ATP1A3;ATP1A4;ATP1B3;ATP2A2;OPRD1;PIK3CG;PPP1CA;PPP1CC;SLC25A6 | ATP1B1;ATP1B4;CALM2;FXYD2;KCNMB2;PPP3R1;RHOA;VDAC2 |
| hsa05210 | Colorectal cancer | 0.02 | AKT2;APC;MSH6;MTOR;PIK3CA;PIK3CB;SMAD4;TCF7 | HRAS;KRAS;MYC;RALA;RHOA |
| hsa05221 | Acute myeloid leukemia | 0.02 | AKT2;CCNA1;JUP;MTOR;PIK3CA;PIK3CB;TCF7;ZBTB16 | HRAS;KRAS;MYC |
| hsa05225 | Hepatocellular carcinoma | 0.02 | AKT2;APC;E2F3;FZD7;IGF1R;MTOR;PIK3CA;PIK3CB;SMAD4;SMARCA2;SMARCD2;TCF7;WNT1;WNT3 | FRAT2;GSTA3;HRAS;KRAS;MYC;SHC1 |
| hsa04664 | Fc epsilon RI signaling pathway | 0.02 | AKT2;BTK;PIK3CA;PIK3CB;VAV3 | HRAS;JMJD7-PLA2G4B;KRAS;MAP2K3;MAP2K4;MS4A2 |
| hsa05226 | Gastric cancer | 0.02 | AKT2;APC;E2F3;FZD7;JUP;MTOR;PIK3CA;PIK3CB;SMAD4;TCF7;WNT1;WNT3 | CDX2;FRAT2;HRAS;KRAS;MYC;SHC1 |
| hsa05206 | MicroRNAs in cancer | 0.03 | APC;CDC25A;E2F3;ERBB3;FOXP1;MTOR;PAK4;PIK3CA;WNT3 | BMF;CRKL;HRAS;KRAS;MYC;RHOA;RPS6KA5;SHC1;STMN1 |
| hsa04978 | Mineral absorption | 0.03 | ATP1A2;ATP1A3;ATP1A4;ATP1B3;SLC5A1 | ATP1B1;ATP1B4;FTL;FXYD2 |
| hsa04625 | C-type lectin receptor signaling pathway | 0.03 | AKT2;NLRP3;PIK3CA;PIK3CB;PLK3;STAT1 | CALM2;CASP8;CLEC7A;HRAS;KRAS;LSP1;PPP3R1;RHOA |
| hsa05152 | Tuberculosis | 0.03 | AKT2;CAMK2G;CEBPB;CIITA; | ATP6AP1;BID;CALM2;C |
| | | | PLK3;RFXAP;STAT1 | AMK2B;CASP8;CLEC7A;HLA-DMB;HLA-DQA2;IL18;LSP1;NFYB;PPP3R1;RHOA |
| hsa04918 | Thyroid hormone synthesis | 0.03 | ATP1A2;ATP1A3;ATP1A4;ATP1B3;DUOX2 | ATP1B1;ATP1B4;CANX;FXYD2;GPX7;PRKACB |
| hsa04668 | TNF signaling pathway | 0.04 | AKT2;CEBPB;MAP3K5;MLKL;PIK3CA;PIK3CB;TRAF1 | CASP8;CCL20;EDN1;ICAM1;MAP2K3;MAP2K4;RPS6KA5 |
| hsa04072 | Phospholipase D signaling pathway | 0.04 | AKT2;GRM3;GRM4;MTOR;PIK3CA;PIK3CB;PIK3CG | ARF6;DGKE;HRAS;JMJD7-PLA2G4B;KRAS;MS4A2;PDGFC;RALA;RHOA;SHC1 |
| **hsa04933** | **AGE-RAGE signaling pathway in diabetic complications** | **0.04** | **AKT2;COL4A4;PIK3CA;PIK3CB;SMAD4;STAT1** | **EDN1;HRAS;ICAM1;KRAS;NOX1;VEGFB;VEGFD** |
| hsa05216 | Thyroid cancer | 0.04 | PPARG;TCF7;TPR | HRAS;KRAS;MYC;TPM3 |
| hsa04720 | Long-term potentiation | 0.04 | CAMK2G;PPP1CA;PPP1CC;RPS6KA1 | CALM2;CAMK2B;HRAS;KRAS;PPP3R1;PRKACB |
| hsa04260 | Cardiac muscle contraction | 0.04 | ATP1A2;ATP1A3;ATP1A4;ATP1B3;ATP2A2 | ATP1B1;ATP1B4;COX6C;FXYD2;TPM3;TPM4 |
| hsa04961 | Endocrine and other factor-regulated calcium reabsorption | 0.04 | ATP1A2;ATP1A3;ATP1A4;ATP1B3 | ATP1B1;ATP1B4;FXYD2;PRKACB |
| hsa04910 | Insulin signaling pathway | 0.04 | AKT2;HK2;HK3;MTOR;PIK3CA;PIK3CB;PKLR;PPP1CA;PPP1CC;PYGL | CALM2;CRKL;HRAS;KRAS;PRKACB;SHC1 |
| hsa04217 | Necroptosis | 0.04 | AIFM1;CAMK2G;HMGB1;HSP90AB1;IFNAR2;MLKL;NLRP3;PARP4;PYGL;SLC25A6;STAT1 | BID;CAMK2B;CASP8;FTL;JMJD7-PLA2G4B;TNFRSF10B;VDAC2 |
| hsa04630 | JAK-STAT signaling pathway | 0.04 | AKT2;CSF3R;GHR;IFNAR2;IL4R;LEP;MTOR;PIK3CA;PIK3CB;PTPN2;STAT1 | HRAS;IFNL3;IL2RA;IL2RG;MYC;OSM;PIAS2 |
| hsa01521 | EGFR tyrosine kinase inhibitor resistance | 0.04 | AKT2;ERBB3;IGF1R;MTOR;NRG1;PIK3CA;PIK3CB | HRAS;KRAS;PDGFC;SHC1 |
| hsa04140 | Autophagy | 0.05 | AKT2;ATG16L1;ATG7;DAPK1;HMGB1;IGF1R;MTMR3;MTOR;PIK3CA;PIK3CB;PIK3R4 | HRAS;KRAS;PPP2CA;PRKACB |
| hsa04390 | Hippo signaling pathway | 0.05 | AFP;APC;BMPR1B;FRMD6;FZD7;GLI2;LATS2;PPP1CA;PPP1CC;SMAD4;TCF7;WNT1;WNT3 | MYC;PPP2CA;SMAD7;STK3 |
| hsa04662 | B cell receptor signaling pathway | 0.05 | AKT2;BTK;CR2;PIK3CA;PIK3CB;RASGRP3;VAV3 | HRAS;KRAS;PPP3R1 |
| hsa04216 | Ferroptosis | 0.05 | ACSL4;ATG7;CP;GSS;SLC39A14 | FTL;VDAC2 |
| hsa04071 | Sphingolipid signaling pathway | 0.05 | AKT2;CERS3;CERS6;MAP3K5;OPRD1;PIK3CA;PIK3CB;PPP2R3A | BID;HRAS;KRAS;MS4A2;PPP2CA;RHOA |

## Figure captions

**Figure 1. Drug target classification.** Direct targets are those that bind to the drug and are responsible for the biological effects of the drug. Indirect targets are those that mediate the biological effects of the drug but do not bind to the drug. By interacting with its direct targets or sequentially modulating its indirect targets through cellular signal transduction, the drug plays its pharmacological action, such as changing the expression of some genes or changing some phenotypes.

**Figure 2. The framework of NP-TCMtarget.**

**Figure 3. An illustration of the NP-TCMtarget website. (A)** The 'Homepage' of NP-TCMtarget shows the overview. **(B)**The DETP module for uploading the query drug/disease gene expression data and the DBTP module for uploading the chemical information represented by SMILES. **(C)** The DTMP module enriches the effect targets into their corresponding biological signaling pathways. **(D)** The result data corresponds to the analysis in the DETP, DBTP and DTMP modules. **(E)** The 'User Guide' page details the procedure and the principle of NP-TCMtarget.

**Figure 4 Exploring molecular mechanisms of XKA in treating type 2 diabetes by NP-TCMtarget. (A)** The analysis process of XKA in treating type 2 diabetes by NP-TCMtarget. **(B)** Pearson correlation of NETS values between XKA and the disease model. **(C)** Pearson correlation of NETS values between XKA and the disease model based on the 734 effect targets that have the potential to mediate the treatment of XKA for the disease model. **(D)** Action path of "active components-direct targets-indirect targets" for XKA in treating type 2 diabetes in the pathway of AGE-RAGE signaling pathway in diabetic complications.